\documentstyle[aps,prl,multicol,psfig]{revtex}

\begin{document}
\draft
\title{Branching Transition of a Directed Polymer in Random Medium}
\author{Giovanni Sartoni$^{1}$\cite{emailG} 
 and Attilio L. Stella$^2$}
\address{$^1$Dipartimento di Fisica and Sezione INFN, Universit\`a di
Bologna, I-40126 Bologna, Italy}

\address{$^2$INFM-Dipartimento di Fisica and Sezione INFN, Universit\`a 
di Padova, I-35131 Padova, Italy}
\date{\today}
\maketitle
\begin{abstract}
A directed polymer is allowed to branch, with configurations determined by 
global energy optimization and disorder. A finite size scaling 
analysis in $2D$ shows that, if disorder makes branching more and more
favorable, 
a critical transition occurs
from the linear scaling regime first studied by Huse and Henley 
[Phys. Rev. Lett. {\bf 54}, 2708 (1985)] to a fully branched, compact 
one. At criticality clear evidence is obtained that the 
polymer branches  
at all scales with dimension ${\bar d}_c$ and roughness exponent $\zeta_c$ 
satisfying $({\bar d}_c -1)/\zeta_c = 0.13\pm0.01$, and energy fluctuation 
exponent $\omega_c = 0.26\pm0.02$, in terms of longitudinal distance.
\end{abstract}
\pacs{Pacs numbers: 05.70.Jk, 64.60.Ak, 61.41.+e, 61.43.-j}
\begin{multicols}{2}  \narrowtext

Linear directed polymers in random medium (DPRM) are a paradigmatic model
 \cite{KardZhang,KPZ},
of central importance for fields like equilibrium interfaces in random 
ferromagnets \cite{Huse} or fracture \cite{Fracture}, in $2D$, and flux lines of high--$T_c$ 
superconductors \cite{Supercond}, in $3D$. The behavior of DPRM at large scales is
generally controlled by properties at temperature $T=0$, where the model
reduces to an optimization problem \cite{KardZhang,Huse}. Imagine we assign independently a
random energy ${\cal E}_b$ to each edge $b$ of a square lattice. At $T=0$,
the optimal DPRM is the linear directed (no overhangs) path $\Pi(t)$, which
covers a given distance $t$ parallel to its direction, e.g. $(1,1)$, and
minimizes $E_{\Pi}=\sum_{b\in \Pi}{\cal E}_b$. Optimization 
determines
a self--affine geometry of the optimal $\Pi$ and a peculiar scaling
of $E_{\Pi}$ fluctuations, quite universal with respect to 
different forms of the environmental disorder\cite{KardZhang}.

So far, much activity on DPRM concentrated on a transition occurring
in high enough $D$ between a low--$T$ regime with rugged free energy
landscape, and an high--$T$ one, with smooth phase space
 \cite{gollinelli,KardZhang}.
In the present Letter we discuss a novel critical transition, triggered by
disorder alone, and occurring in
an appropriate generalization of DPRM. We release the constraint
of linearity and consider polymers which can form branches and loops.
At the same time, we assume that the random energy ${\cal E}_b$ can take both
positive and negative values. We look for the branched directed polymer
$\Pi$ (BDPRM) spanning a certain longitudinal distance $t$ and optimizing 
$E_{\Pi}$ ( defined as in the linear case). Even a very little percentage
of negative energy bonds suffices now to induce branches and loops
in the optimal $\Pi$, at least at small scales. Indeed, a single negative 
bond met along 
an optimal linear path can easily become a branch, if it lowers
$E_{\Pi}$ once included in $\Pi$.
However, as long as the weight of negative energies in the 
${\cal E}_b$--distribution
remains low, the optimal $\Pi$ is likely to remain linear at large scales.
On the other hand, if this weight is very large, one expects branches and loops
to develop wherever possible, leading to a sort of fully ramified, compact
structure for $\Pi$.

For a specific model, we find here that 
such two limiting regimes hold on respective sides
of a sharp threshold concentration of negative energy bonds. Right at threshold,
a critical transition occurs, with $\Pi$ characterized by
a branching probability between $0$ and $1$
at all scales. This critical regime, with an infinite hierarchy
of loops and dangling ends, is the first of this kind 
met in this field and demonstrates an unexpected richness of possible 
solutions of problems of global optimization with disorder. The transition 
has analogies with theta-- or similar points in polymer physics. 

A main motivation to study transitions like this comes from recent attempts
to explain the puzzling universality of $2D$ random Potts  
ferromagnets \cite{frothM}.
To the purpose of evaluating its free energy, the interface between two phases
of a $q$-state Potts model can be schematized as a directed path, which possibly
includes bubbles of the other $q-2$ phases \cite{frothM}. At $T=0$ these bubbles are
induced by the presence of antiferromagnetic couplings, which naturally
proliferate under renormalization. Such interfaces with bubbles are ramified
polymers without dangling ends, and obey global optimization. 
In the ferromagnetic phase bubbles should not develop at large scales,
while in the disordered phase they should form a compact "froth" structure.
The existence
and nature of peculiar, bubbling critical regimes for these polymers
can be an important element to understand the above universality \cite{frothM}.
Another potential field of application are models of fracture in disordered 
materials \cite{Fracture}.

Counting the configurations of a ramified structure, even if directed,
is a problem of non--polynomial complexity. An additional difficulty hindering
progress in these problems comes
from the need of sampling adequately over disorder.
Thus, our study here is based on a quite extensive 
computational effort, combined with a suitable choice of model and method of 
scaling analysis. We consider bond branched polymers directed at $45^\circ$  
with respect to the coordinate directions of a square lattice, with origin 
in $O$ (Fig.1a). 
No branch can turn backwards with respect to the polymer direction.
A polymer, $\Pi (t)$,
extends from $O$ up to the straight line, $S_t$, at 
distance $t$ from $O$ and orthogonal to the polymer direction (Fig. 1a). 
For bond energies we choose a distribution $\displaystyle P({\cal E}_{b})=
p\delta({\cal E}_{b}-1)+(1-p)\delta(
{\cal E}_{b}+1)$ with $0\leq p\leq 1$\cite{nota}. 
Upon decreasing $p$ from $1$, 
we expect to possibly meet a threshold as described above at some
$p=p_c$.
\begin{figure}
\centerline{
\psfig{file=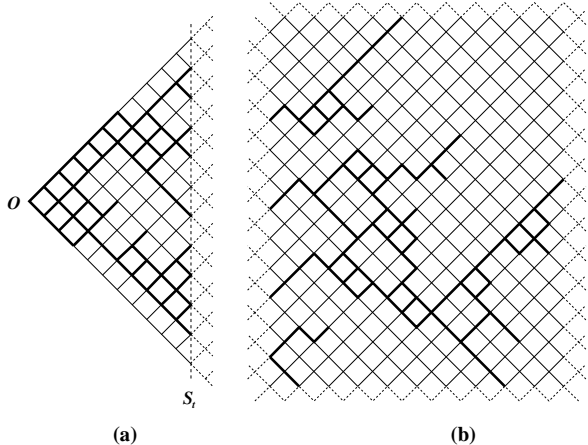,height=5.7cm}}
\vskip 0.2truecm
\caption{(a) BDPRM configuration without strip restriction. $t$ is the number 
of steps of any linear path joining $O$ with $S_t$. (b) A portion ($
\Delta t=21$) of optimal BDPRM confined within a strip 
($L=13$). Disorder is generated with $p=0.925$. 
Strips are periodic in the $L$ direction.}
\end{figure}
Information on the optimal path in a given disorder configuration
is transferred  from $t$ to $t+1$ by updating 
the set $\{E(\tau,t)\}$ of
the optimal $\Pi$ energies for given $t$ and $\tau=\Pi(t)\cap S_t$. 
The recursion for $E$ reads:
\begin{equation}\label{energyTM}
E(\tau',t+1)=\min_{\tau,\sigma}[E(\tau,t)+\epsilon(\sigma)]
\end{equation}
where $\tau'=\Pi(t+1)\cap S_{t+1}$ and $\epsilon(\sigma)$ is the energy 
associated to each admissible set of
bonds $\sigma(\tau,\tau')$ connecting points in $\tau$ to points in $\tau'$.
There are typically several  such sets for given $\tau$ and $\tau'$. The least
energy, $E_{\mbox{\scriptsize min}}(t)=\displaystyle\min_{\Pi(t)} E_{\Pi(t)}$
is then calculated as $E_{\mbox{\scriptsize min}}(t)=\displaystyle\min_\tau
E(\tau,t)$. Thus, in order to identify the optimal $\Pi$, at any given  $t$,  
we have to determine $E(\tau',t')$ for all $t'<t$ and $\tau'\subset  S_{t'}$, 
consistently with eq.(\ref{energyTM}). There are $2^{t+1}-1$ different $\tau$
intersections corresponding to all non-empty subsets of points in $S_t$, and
altogether $2(t+1)$ bonds connecting $S_t$ to $S_{t+1}$, hence 
a transfer algorithm 
must scan $2^{2t+2}-1$ $\sigma$-configurations to calculate the minimum
in (\ref{energyTM}). This means in total $2^{3(t+1)}$ 
operations per $t$-step. Thus, as soon as $t$
exceeds a few units, a reasonable statistics over disorder can not be 
collected. Clearly, one cannot simply apply the strategy of following the polymer
development within the triangle in Fig.1a up to large $t$:
a different method of scaling analysis and some restrictions on allowed 
configurations are in order. Thus, we confine 
$\Pi$ within a strip of fixed width, $L$, and length $t$ 
(Fig.1b). The 
$t$-transfer of information on optimal configurations within such a 
strip has a $t$--independent computational cost determined by $L$. 
Our strategy is to extract
information on asymptotic properties from strip calculations for relatively small
$L$ by exploiting finite size scaling \cite{Seno}.
      
Since transfer can be made much 
faster by restricting to only one $\sigma$ the possible connections between 
two compatible $\tau$ and $\tau'$ in eq.(\ref{energyTM}), the following 
constraint 
is applied to determine such unique $\sigma$: BDPRM configurations are  
assumed to be always such to necessarily imply bond connection between 
neighboring sites in $\tau\subset S_{t}$ and in in $\tau'\subset S_{t+1}$.
%
%
Namely the unique $\sigma$ considered in (\ref{energyTM}) is that 
accomplishing all possible nearest neighbor connections between the considered $\tau$ and $\tau'$.
Such restriction offers the advantage of reducing the operations necessary 
to determine $E(\tau',t+1)$ in (\ref{energyTM}) by a factor 
$2\times2^L$. On the other hand, such $\sigma$ represents the contribution 
to transfer matrix elements with the highest rate of loop and branch formation, leaving 
also a wide  possibility of dangling ends in the structure, and thus making 
the model fully adequate to display the transition we are interested in. 
Indeed, we also performed extensive calculations
for an ordered version of this restricted model
at $T>0$ to test the effects of such restriction. 
On strips with $L$ up to 10, with methods similar to those in use for
 directed lattice animals (DA) \cite{Derrida},
we estimated correlation length exponents 
$\nu_{\scriptscriptstyle \|}=0.81\pm0.02$ and $\nu_{
\scriptscriptstyle\bot}=0.48\pm0.02$,
to be compared with
$\nu_{\scriptscriptstyle \|}=0.818\pm 0.001$ 
and $\nu_{\scriptscriptstyle \bot}=1/2$ for DA \cite{Derrida,exactDA}. 
This suggests that our model, without disorder, belongs to the same 
universality class as unrestricted DA. 
  
We identify the transverse BDPRM width, at a given $t$, as the root mean
square distance of the points 
in $\tau$ from the middle of the strip. Since 
we expect such width, averaged over random bond configurations, 
to grow like $t^\zeta$ ($0\leq\zeta\leq 1$) on infinite 
lattice, crossover on our strips should be controlled by the ratio $t^\zeta /L$.
So, the average total number of bonds should scale with $L$ and $t$ as:
\begin{equation}\label{masscal}
{\cal N}(L,t)\equiv\langle N(L,t)\rangle=t^{\bar d}n\left({t^\zeta 
\over L}\right)
\end{equation}
where $\bar d$ is a fractal dimension and $n$ is a crossover function . The 
brackets indicate average over disorder. For $t\:,\: L\gg1$, and 
$t^\zeta \ll L$ 
we should recover the behavior on unrestricted lattice. So $\displaystyle 
\lim_{x\rightarrow 
0}n(x)=const.\not= 0$. On the other hand, for $t^\zeta \gg L$, $\cal N$ must 
become proportional to $t$ as a consequence
of statistical invariance under $t$-translations. Thus, $\displaystyle 
n(x)\sim_{\scriptscriptstyle x\rightarrow\infty}
x^{1-{\bar d}\over\zeta}$. This means that $\displaystyle {\cal N}(L,t)\sim_{
\scriptscriptstyle t\gg L}
A(L)t$, with $A(L)\sim L^{({\bar d}-1)/\zeta}$.
In a similar way, for the fluctuations of optimal path 
energies due to lattice disorder we expect
\begin{equation}\label{sigenscal}
\Delta{\cal E}(L,t)\equiv\langle \left( E_{\mbox{\scriptsize min}}(t) -
\langle E_{\mbox{\scriptsize min}}(t) \rangle\right)^2 
 \rangle^{1/2} \sim t^\omega e\left({t^\zeta \over L}\right)
\end{equation}
For $t^\zeta \gg L$, $\Delta {\cal E}$ should grow 
like $t^{1/2}$, 
as a consequence of the central limit theorem \cite{CentLiTh}. The amplitude of 
this  $t^{1/2}$ behavior should scale as $\displaystyle B(L)\sim 
L^{(\omega -1/2)/\zeta}$.
On the other hand, for $t^\zeta \ll L$, $e\sim const.$, as $\omega$ 
describes $\Delta{\cal E}$ scaling without strip restriction.
Of course, functions like $n$ and $A$, and the various exponents,
can depend on $p$.
   
We explored systematically the scalings in eqs.(\ref{masscal}) and 
(\ref{sigenscal}) for different $p$'s. $L$ and $t$ ranged up to $9$ and several 
hundreds, respectively, and data for $\cal N$ and $\Delta {\cal E}$ 
resulted from 
statistics over a minimum of $2\times10^3$ up to a maximum of $10^5$
disorder configurations. The most 
convincing  exponent determinations always occurred when 
analyzing regimes at $t^\zeta \gg L$. E.g., the amplitude $A(L,p)$
can always be estimated with good accuracy (relative 
error bars $<10^{-2}$), being determined by averaging over large data sets. 
The trend of the effective exponents $\displaystyle \ln\left[ A(L+1,p)/A(L,p)
\right]/\ln
[(L+1)/L]$ can be extrapolated rather well for  all $p$'s (Fig.2).
For $p\lesssim 0.90$ extrapolation to $({\bar d}-1)/\zeta\simeq1$ is 
always clear. This indicates a compact, fully branched regime
(${\bar d}=2,\zeta=1$). On the 
other hand, for $0.94\lesssim p\lesssim 0.98$, we extrapolate $({\bar d}-1)/
\zeta\simeq 0$ \cite{peq1}. This suggests ${\bar d}=1$, i.e. a 
linear regime.
\begin{figure}
\centerline{
\psfig{file=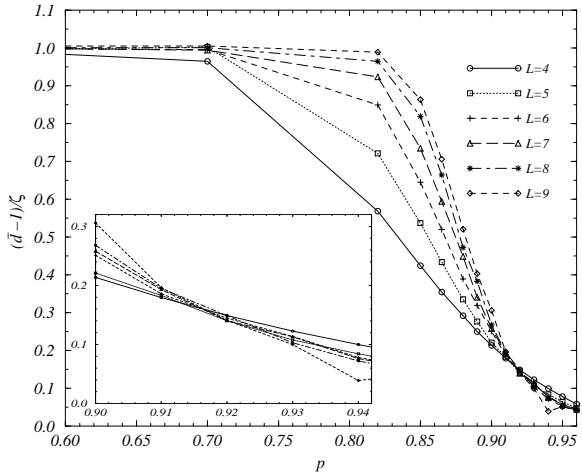,height=7.3cm}}
\caption{Effective determinations of $({\bar d}-1)/\zeta$ for 
different $L$ and varying $p$. 
The insert shows a magnification of the transition region, where exponent 
curves intersect.}
\end{figure}
Most interesting, there is a narrow region ($p\sim 0.92\div 0.93
$) at which the trends of approach to the above two asymptotic values exchange 
each other, and intersections of effective exponent ``curves'' for 
different $L$
concentrate and stabilize. The corresponding value $({\bar d}_c-1)/
\zeta_c \simeq0.12\div 0.14$ is an obvious candidate to represent a distinct 
border regime right at $p=p_c \simeq0.925$. The data in Fig.2 suggest a step 
at $p=p_c$ for $L\rightarrow\infty$. The scenario is further 
elucidated by a similar analysis of $B$ amplitudes in eq.(\ref{sigenscal}). 
This time we get clearly $(\omega -1/2)/\zeta\simeq 0.50$ for 
$p\lesssim 0.86$ and 
$(\omega -1/2)/\zeta\simeq -0.27$ for 
$0.94\lesssim p \lesssim 0.98$. The value $0.50$ 
confirms a compact, fully branched regime, for which we 
expect $\zeta=1$, and also, rather naturally, $\omega=1$\cite{CentLiTh2}.
$(\omega -1/2)/\zeta\simeq -0.27$ in the $\bar d \simeq 1$ region is clear 
evidence of standard, linear DPRM scaling ($\omega=1/3\:,\:\zeta=2/3$ \cite
{KardZhang,Huse}). Thus, in this regime the presence of branchings at
small scales does not change the universal behavior with respect to the
strictly linear case.
Unfortunately, it is more difficult to guess precisely the border-line 
value of $(\omega_c -1/2)/\zeta_c$. This is 
probably due to the circumstance that for $p=
p_c$ this combination of exponents seems to be lower 
than in the regimes on both sides. So, the crossover pattern can not stabilize 
simply and
nicely as in the case of Fig.2. A direct determination of $\omega$ in the short 
time regime ($t\lesssim L=9$), by fitting eq.(\ref{sigenscal}) directly,
 gives $\omega_c =0.26\pm0.02$, 
distinct from the DPRM $\omega =1/3$ \cite{DPRMom}. Assuming $\zeta_c$ equal 
to $1$, this $\omega_c$ is consistent with our extrapolation
of $(\omega_c -1/2)/\zeta_c$ from $B(L,p_c)$ ($\omega_c=0.27\pm 0.02$). 
Further insight comes from data collapse 
plots testing $A$ in the form
\begin{equation}\label{collapse}
A(L,p)=L^{({\bar d}_c -1)/\zeta_c} a\left( (p-p_c)L^\phi \right)\ \ ,
\end{equation}
where $\phi$ is a crossover exponent (Fig.3). Eq.(\ref{collapse}) 
follows from (\ref{masscal}), once explicitated the dependence on the relevant 
parameter (in renormalization group (RG) sense),  $p-p_c$. 
Over the range studied, data collapse best for $p_c =0.927\pm 0.005$,
$({\bar d}_c -1)/\zeta_c =0.13\pm0.01$ 
and $\phi =1.00\pm 0.02$. $\displaystyle a(x)={\cal A} e^{-\alpha x}$ seems to 
be an acceptable form for the crossover function $a$. 
%
%
\begin{figure}
\centerline{
\psfig{file=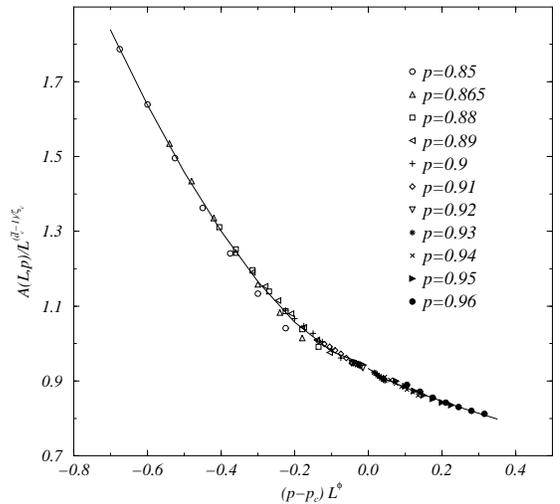,height=7.3cm}}
\caption{Collapse plot of the rescaled mass-amplitude data.}
\end{figure}
The above analysis shows that, upon increasing the fraction 
of negative bond energies on the lattice, BDPRM undergo a sharp, critical 
transition from linear DPRM, to compact, fully branched regimes. At the 
transition $({\bar d}_c -1)/\zeta_c =0.13\pm0.01$, $\omega_c =0.26\pm 0.02$
and $\phi=1.00\pm 0.02$,
indicating a fractal regime, with branches on all scales,
and peculiar energy fluctuations exponent. 
For sure ${\bar d}_c$, $\zeta_c$ are different from 
the corresponding exponents of DP (${\bar d}=1.473\pm0.001$ \cite{Roux}, 
$\zeta=0.6330\pm0.0008$ \cite{EssamDP}, $({\bar d}-1)/\zeta=0.747\pm0.002$), 
which is a reference model for a variety 
of phenomena, including fracture \cite{DP,DPdisorder}. Our critical regime appears
a peculiar result of optimization and disorder.
    
It is natural to ask whether the $T=0$ behavior analyzed above should 
apply also at $T>0$.
A numerical investigation of $T>0$ properties in the random case would be 
computationally
too expensive, while higher dimensions
are definitely beyond present possibilities.
For $p=1$ and $T>0$ our model describes DA
with bond fugacity $\exp(-1/T)$. Application of 
Harris' criterion \cite{Harriscrit} to the 
DA critical point ($T_c=0.9708\ldots$) suggests that disorder
is relevant for our model in $2D$, because $\displaystyle 
\nu_{\scriptscriptstyle \|}
[1+(d-1)\zeta]<2$ (see exponents quoted above). We expect a
critical line joining the critical points $(1,T_c)$ and $(p_c,0)$ on
the $(p,T)$ plane.
The point $(p_c,0)$ should be attractive from an RG 
point of view. Indeed, in a coarse-grained description, based on a
branched path integral, for the BDPRM at $T>0$ one would expect an elastic free
energy term $ {\scriptstyle-{1\over T}} \int \left({dx_i \over dt}
\right)^{\scriptscriptstyle2} {\scriptstyle dt}$ associated 
with the i-th branch. This implies 
${dT/ d\varepsilon}= (1-2{\bar\zeta})T$ if $t$ is rescaled by a 
factor $1+\delta\varepsilon$ and $\bar{\zeta}$ is the
roughness of a single branch.  If $\bar\zeta >1/2$, which 
is plausible in our case, irrelevance in $T$ follows.
So, the whole critical line for $p<1$ could be
controlled by the $T=0$ fixed point tested by our scaling analysis.
In higher $D$, while disorder should remain relevant on the basis of Harris'
criterion\cite{Day}, the transition of DPRM to smooth high--$T$ 
free energy landscape \cite{gollinelli} could modify the above phase diagram.
  
Summarizing, we gave evidence that properly chosen disorder conditions
determine a branched critical regime in a generalized version
of DPRM. In a previous study, a similar model without dangling ends on 
hierarchical lattice, displayed a strictly linear (${\bar d}_c=1$) $T=0$  critical regime \cite{frothM}.
%
%
So, dangling ends
could be essential for obtaining ${\bar d}_c>1$ in a BDPRM model.
Further clarification of this issue may be a crucial test for
the scenario drawn in ref.\cite{frothM}, concerning the universality mechanism
in random ferromagnets with several coexisting phases. However, a considerably
more substantial computational effort would be required to such purpose.
A comparison with DP clusters is also appropriate for our critical branched
polymer. It was shown recently that the DP cluster backbone is the structure
hosting and shaping (in the sense of Hurst's exponent) optimal 
linear paths for an extremal version of the DPRM model \cite{RZ}.
The branched fractal cluster identified here, is clearly different from the DP
one, and originates in a complete and direct way from 
global optimization and disorder. 
     
We thank Mehran Kardar for useful comments.

\end{multicols}

\begin{references}
\bibitem{nota}We got numerical indications that different forms
of $P$ would not change our main results.
\bibitem{CentLiTh}In this limit $\Pi$ fills an effectively $1D$ 
lattice, and $E_{\Pi}$ is the sum of many independent 
contributions.
\bibitem{peq1}For $p=1$, the model is ordered. So, crossover to
$T=0$ directed polymer behavior ($\omega=0$, $\zeta=1/2$) occurs for
$p\approx1$. In spite of this, the window for detecting DPRM 
behavior with $ p_c<p< 1$ is sufficiently wide.
\bibitem{CentLiTh2}Upon doubling its linear size, the compact structure 
grows into the union of $4$ of the original units. Assuming weak dependence of 
the random energies 
of such units leads to $\omega=1$.
\bibitem{DPRMom}
Similar calculations for $0.94<p<0.97$, i.e. within the linear DPRM regime,
give $\omega=0.33\pm0.02$, which is consistent with the expected value. This
suggests to trust our extrapolation of short $t$ data.
\bibitem{Roux}The fractal dimension of the critical 
DP cluster can be calculated as ${\bar d}=1+\zeta-(\beta/
\nu_{\scriptscriptstyle \|})$, along lines illustrated, e.g., by 
Roux and Zhang in ref.\cite{RZ}.
\end{references}
\end{document}